\documentclass[letterpaper,titlepage,11pt]{article}
\pdfoutput=1
\usepackage{hyperref}

\usepackage{amssymb,amsmath,amsfonts}
\usepackage{epsfig}
\usepackage{array}
\usepackage{booktabs}

\def\clock{{\count0=\time
          \divide\count0 60
          \ifnum\count0<10 0\fi\the\count0
          \multiply\count0 -60 \advance\count0 \time
          :\ifnum\count0<10 0\fi \the\count0
        }}
\newcommand{\timestamp}{{\small\vbox{\hbox{\tt\jobname.tex}
\hbox{\the\day/\the\month/\the\year, \clock}}}}

\newcommand{\CF}{\mathcal{F}}

\newcommand{\CL}{\mathcal{L}}
\newcommand{\CO}{\mathcal{O}}
\newcommand{\CN}{\mathcal{N}}

\newcommand{\CR}{\mathcal{R}}

\newcommand{\gym}{g_{\rm YM}}

\newcommand{\ads}{\mbox{AdS}}

\newcommand{\nn}{\nonumber}

\newcommand{\spa}{\ , \ \ }

\newcommand{\beq}{\begin{equation}}
\newcommand{\eeq}{\end{equation}}

\newcommand{\ds}{\displaystyle}

\newcommand{\tr}{\mathop{{\rm Tr}}}

\newcommand{\hod}{ {}^* \!}

\numberwithin{equation}{section}

\begin{document}

\begin{titlepage}
\ \ \vskip 1.8cm
\centerline {\huge \bf The Born-Infeld/Gravity Correspondence} 
\vskip 1.4cm

\centerline{\large {\bf Gianluca Grignani$\,^{1}$},  {\bf Troels Harmark$\,^{2}$},}
\vskip 0.2cm \centerline{\large  {\bf Andrea Marini$\,^{1}$}
and
{\bf Marta Orselli$\,^{1,2}$} }

\vskip 0.7cm

\begin{center}
\sl $^1$ Dipartimento di Fisica, Universit\`a di Perugia,\\
I.N.F.N. Sezione di Perugia,\\
Via Pascoli, I-06123 Perugia, Italy
\vskip 0.3cm
\sl $^2$ The Niels Bohr Institute, Copenhagen University  \\
\sl  Blegdamsvej 17, DK-2100 Copenhagen \O , Denmark
\end{center}
\vskip 0.3cm

\centerline{\small\tt grignani@pg.infn.it, harmark@nbi.dk, }
\centerline{\small\tt andrea.marini@fisica.unipg.it, marta.orselli@unipg.it}

\vskip 1.2cm \centerline{\bf Abstract} \vskip 0.2cm \noindent
In this paper we explore the correspondence between four-dimensional Born-Infeld theory and five-dimensional classical gravity. The Born-Infeld theory side corresponds to the low energy effective theory for open strings ending on coincident D3-branes in a (slowly varying) background Kalb-Ramond field, including all higher-derivative corrections. On the gravity side one has the gravitational (closed string) description of D3-branes in the same background Kalb-Ramond field and the correspondence is thus a consequence of the open/closed string duality. According to the correspondence the gravity side provides a description of the strong coupling limit of Born-Infeld theory. This is a correspondence between effective theories in a similar sense as in the fluid/gravity correspondence. We match the Born-Infeld and gravity sides up to, and including, two-derivative corrections. To this end, we find a new gravity solution for D3-branes with flat embedding in the background of an arbitrary constant background Kalb-Ramond field and show that there are no two-derivative corrections to this for a slowly varying Kalb-Ramond field.

\end{titlepage}

\tableofcontents
\pagestyle{plain}
\setcounter{page}{1}

\section{Introduction}
\label{sec:intro}

In this paper we explore a correspondence between Born-Infeld theory and gravity. On the one side, we have weakly coupled four-dimensional Born-Infeld theory including higher-derivative corrections. This is the low energy effective theory for open strings ending on $N$ coincident D3-branes with flat embedding and a slowly varying Kalb-Ramond field in the background. On the other side, corresponding to the strong coupling limit of the Born-Infeld theory, we have a solution of five-dimensional classical gravity. This is the low energy effective gravitational description of $N$ coincident D3-branes with flat embedding in the same Kalb-Ramond field background.

The Born-Infeld/gravity correspondence can be seen as a holographic correspondence since it relates the strong coupling limit of a theory without gravity to a gravitational theory. Since it is a correspondence between effective theories it is akin to the fluid/gravity correspondence \cite{Bhattacharyya:2008jc,Hubeny:2011hd}. However, unlike the fluid/gravity duality, the duality considered here is at zero temperature and the gravitational side involves asymptotically flat solutions of gravity rather than asymptotically AdS. 

For the open-string (Born-Infeld) side at weak coupling the leading order effective action is~\cite{Fradkin:1985qd,Abouelsaood:1986gd}
\begin{equation}
\label{introBI}
I = -  N T_{\rm D3} \int d^4\sigma \sqrt{- \det ( \eta_{ab} + \CF_{ab} ) }
\end{equation}
This is the effective action for $N$ D3-branes in the background of ten-dimensional flat space with a slowly varying Kalb-Ramond field $B_{ab}$ and with the Ramond-Ramond fields being zero. We specialize here to the case of zero world-volume field strength which means that the action effectively is the abelian Born-Infeld action times $N$ and that the abelian part of the gauge-invariant field strength is $\CF_{ab} = B_{ab}$ in the world-volume directions. The above action is computed (for $N=1$) from the disc topology of the open string world-sheet. One has two possible expansion directions. Either in terms of the $\alpha' = l_s^2$ parameter, $l_s$ being the string length. Or in terms of the topological expansion of the string world-sheet. 

For a varying $\CF_{ab}$ the action \eqref{introBI} will have higher-derivative $\alpha'$ corrections. From covariance it follows that there are only corrections at an even number of derivatives. Each derivative comes with a string length $l_s$ thus this expansion is in the low energy limit $l_s \partial_a \CF_{bc} \ll 1$. It has been shown that there are no two-derivative corrections \cite{Andreev:1988cb}. Moreover, all terms with $2n$ derivatives and $n+2$ powers of $\CF_{ab}$ have been found \cite{Andreev:1988cb,Wyllard:2000qe,deRoo:2003xv}.

Turning on the string coupling $g_s$ one needs to consider the topological string world-sheet expansion. We explain that for many branes $N \gg 1$ this can be separated into planar and non-planar diagrams, with the planar diagrams being dominant. The appropriate coupling to expand in is $g_s N$ which is analogous to the 't Hooft coupling of Yang-Mills theories. The power of $g_s N$ is set by the number of boundaries of the world-sheet. Hence when $g_s N$ becomes finite or large one needs to add diagrams with an arbitrary number of boundaries. 

Low energy description of D-branes can either be described in terms of open string variables or closed string variables due to the open/closed string duality \cite{Polchinski:1995mt}. The low energy description of the closed string side is type IIB supergravity. When describing $N$ coincident D3-branes using supergravity solutions one has included string diagrams with an arbitrary number of boundaries since the supergravity solution is non-linear. Therefore, on the closed string side it is possible to describe the $g_s N \gg 1$ regime. In fact, it is a necessary condition for the validity of the supergravity approximation.

In this way one gets a Born-Infeld/gravity corrrespondence as a low energy manifestation of Polchinski's discovery of the open/closed string duality, namely that D-branes not only are what open strings end on, but that they also source the closed string Ramond-Ramond fields \cite{Polchinski:1995mt}. Considering $N \gg 1$ coincident D3-branes in a slowly varying Kalb-Ramond field, then in the weakly coupled regime $g_s N \ll 1$ one has a good description in terms of the low energy effective theory for open strings, described by the Born-Infeld action plus higher derivative corrections from the $\alpha'$ expansion. Instead in the strongly coupled regime $g_s N \gg 1$ one has a good description in terms of a five-dimensional gravity theory, dimensionally reduced from type IIB supergravity. In this case, the leading higher-derivative corrections come in powers of $\sqrt{g_s N} \alpha'$.

Our first main result of this paper is that we formulate the higher-derivative expansion on the gravity side, and we show how to read off both the equations of motion for the Born-Infeld field strength as well as the energy-momentum tensor. 

The second main result is the zeroth order gravity solution, presenting for the first time the general solution of $N$ coincident D3-branes in the background of a constant Kalb-Ramond field. This is presented in a covariant form with respect to the relativistic symmetry on the world-volume. In this form the solution is actually simpler than the solutions it generalizes, such as the $D3 \parallel D1$, $D3 \parallel F1$ \cite{Lu:1999uca}, $D3 \parallel (F1 \parallel D1)$ and $D3 \parallel (F1 \perp D1)$ bound states \cite{Green:1996vh,Russo:1996if,Breckenridge:1996tt,Costa:1996zd,Cederwall:1998tr,Lu:1999uv,Grignani:2013ewa}. From this we reproduce the leading order Born-Infeld action \eqref{introBI} in the strongly coupled regime $g_s N \gg 1$.

The third main result is that we can show that the two-derivative correction on the gravity side does not give rise neither to a two-derivative correction of the energy-momentum tensor, nor to the equations of motion of the field strength. This result rests on showing that the zeroth order solution is unique under small perturbations, which we are able to show using among other things a chain of T-dualities. Thus, we deduce from this that the Born-Infeld action at strong coupling $g_s N \gg 1$ does not have two-derivative terms at order $\sqrt{g_s N} \alpha'$. This seems to be in accordance with the absence of two-derivative terms for the weakly coupled regime $g_s N \ll 1$ at order $\alpha'$. Thus, this result points to a possible connection between the $\alpha'$ expansion at weak coupling $g_s N \ll 1$ and the $\sqrt{g_s N} \alpha'$ expansion at strong coupling $g_s N \gg 1$.

The Born-Infeld/gravity correspondence is closely related to the blackfold effective theory approach to branes \cite{Emparan:2009cs,Emparan:2009at}.%
\footnote{See \cite{Caldarelli:2010xz,Grignani:2010xm,Emparan:2011hg,Armas:2011uf,Camps:2012hw,Armas:2013hsa,Armas:2014rva} for further developments.} The blackfold approach enables one to make an effective description of branes at finite temperature in a situation where the thickness of the brane is much smaller than the length scales associated with its embedding and the background of the brane. The effective theory is described as a fluid living on the brane. Indeed, in \cite{Emparan:2013ila} it is shown how one can interpolate between the blackfold approach for D3-branes and the fluid/gravity duality. Specifically for D-branes, it has been pointed out previously that the blackfold approach is in correspondence with the Dirac-Born-Infeld action \cite{Grignani:2010xm,EmparanUnpublished,Grignani:2013ewa,Niarchos:2014maa}. Earlier work on this type of correspondence can be found in \cite{Emparan:2001ux,Lunin:2007mj,Lunin:2008tf}. In this paper we make a more precise study of this type of correspondence by considering a specific setting where the world-volume has a flat embedding.

While in this paper we consider Born-Infeld theory at zero temperature, one can also explore the Born-Infeld/gravity duality at non-zero temperature. The first step towards this was taken in \cite{Grignani:2013ewa} where one compares the thermal Born-Infeld theory with a constant field strength at weak and strong coupling. The main result of this is that the leading $T^4$ correction to the free energy at strong and weak coupling only differ by the famous $3/4$ factor of \cite{Gubser:1996de} despite the non-trivial dependence on the field strength. The reason for this is that taking the decoupling limit leads to the AdS/CFT correspondence in the Poincar\' e patch though in a different coordinate system \cite{Grignani:2013ewa}. 

In the literature one has investigated the possibility of a full duality between type IIB string theory on the background of $N$ D3-branes and a theory that would reduce to $\CN=4$ super Yang-Mills theory in the infrared \cite{Gubser:1998kv,Gubser:1998iu,Park:1999xz,Intriligator:1999ai,Danielsson:2000ze,Rastelli:2000xj,Amador:2003ju}. In case one can formulate such a generalized version of the AdS/CFT correspondence the Born-Infeld/gravity correspondence should emerge in the low energy limit. We would like to emphasize that our results are independent on whether this is possible, or not. Indeed, the Born-Infeld/gravity correspondence is between two effective descriptions of the D3-brane physics that both arise considering Polchinski's open/closed string duality \cite{Polchinski:1995mt} in the low energy regime, and as such only relies on the validity of the open/closed string duality. 

This paper is structured as follows. First we explain the general setup with D3-branes in a background Kalb-Ramond field in Section \ref{sec:setup}. Secondly, we have some general remarks about why there is a Born-Infeld/gravity duality in Section \ref{sec:OCduality} and how it arises naturally from the open/closed string duality. Then in Section \ref{sec:BIside} we review the Born-Infeld theory side, including some of the leading higher-derivative corrections. In Section \ref{sec:gravside} we turn to the gravitational side of the duality. First we give the general ansatz for the gravity solution and the equations of motion. Then in Section \ref{sec:zerosol} we present the new general solution for a D3-brane with arbitrary constant Kalb-Ramond field. In Sections \ref{sec:duality_chain}-\ref{sec:uniqD3} we show that this general solution can be generated by a chain of T-dualities, rotations and boosts, and that the solution is unique under small perturbations. In Section \ref{sec:corrected_sol} we consider the systematic expansion of the gravity solution in powers of higher-derivative corrections along the world-volume direction. We show how to read off the equations of motion for the Born-Infeld field strength at an odd number of derivatives and show furthermore how the solution is corrected at two derivatives. Finally, in Section \ref{sec:reading_EM} we explain how to read off the higher-derivative expansion of the energy-momentum tensor for the Born-Infeld theory. We end the paper with our conclusions and a discussion of the results in Section \ref{sec:concl}.

\noindent {\bf Important note:} While this paper was in the final stages of preparation, the paper \cite{Niarchos:2015moa} appeared, with independent work that has significant overlap with this one.

\section{The setup}
\label{sec:setup}

We consider $N$ coinciding D3-branes with $N \gg 1$ in the background of ten-dimensional Minkowski space. Writing $x^\mu$, $\mu=0,1,...,9$, as the Cartesian coordinates for the Minkowski space we put the D3-branes at the hyperplane $x^4=x^5= \cdots = x^9 = 0$. More specifially, writing the world-volume coordinates for the D3-branes as $\sigma^a$, $a=0,1,2,3$, we choose the flat embedding $x^a(\sigma) = \sigma^a$, $a=0,1,2,3$ and $x^i(\sigma)=0$ for $i=4,5,...,9$. 

The type IIB supergravity background furthermore has zero dilaton field $\phi=0$ and zero Ramond-Ramond field strengths. However, we turn on a Kalb-Ramond field $B_{\mu\nu}$ in the directions parallel to the brane world-volume while being zero along the transverse directions. We write this as $B_{ab}=\CF_{ab}$ for $a,b=0,1,2,3$ and $B_{\mu\nu}=0$ otherwise, where $\CF_{ab}$ can be thought of as the induced gauge invariant field strength on the D3-branes.
From the type IIB supergravity equations we get that the Kalb-Ramond field strength is zero $dB_{(2)}=0$ and hence
\begin{equation}
\label{BI_FS}
\partial_{[a}\CF_{bc]}=0
\end{equation}
Due to the presence of the D3-branes this is not pure gauge. 

We assume that $\CF_{ab}$ is slowly varying over the D3-brane world-volume. Let $R$ denote the minimal length scale of variation of $\CF_{ab}$. Then slowly varying means that $R/l_s \gg ( g_s N)^{\frac{1}{4}}$ as well as $R/l_s \gg 1$. This ensures that the variation of $\CF_{ab}$ is over scales much larger than the effective thickness of the D-branes and that we can integrate out the string scale to work with an effective low energy description.

When $g_s N \ll 1$ the open-string dynamics is weakly coupled and hence we use open strings to describe the D-branes. The effective low energy description where we integrate out the massive open string modes is well-known to be the Born-Infeld action to leading order. We consider this in Sections \ref{sec:OCduality} and  \ref{sec:BIside}.

As we discuss in detail below in Section \ref{sec:OCduality}, when $g_s N \gg 1$ the open-string dynamics is strongly coupled but one can instead use the closed string channel to describe the D-brane dynamics, employing Polchinski's open/closed string duality. Here the low energy effective description is type IIB supergravity. The D-branes backreact on the geometry and we can describe their dynamics by finding solutions to type IIB supergravity, which for our setup effectively reduces to finding solutions of a certain five-dimensional theory of gravity with some matter fields. We consider this in Section \ref{sec:gravside}.

\section{General remarks on the Born-Infeld/gravity correspondence}
\label{sec:OCduality}

Before we turn to the low energy effective descriptions of the Born-Infeld/gravity correspondence from open and closed string point of view in Sections \ref{sec:BIside} and \ref{sec:gravside}, we first consider how one can understand it on general grounds in terms of the open/closed string duality. 

\subsubsection*{Open string point of view}

We consider the setup of $N$ D3-branes in a slowly varying background Kalb-Ramond field as described in Section \ref{sec:setup}. Viewing this from the open string point of view one can compute a low energy effective action for the dynamics of the D3-brane using perturbative open string theory. One has two relevant expansion parameters: $\alpha'$ and the string coupling $g_s$. $\alpha' = l_s^2$ is the inverse of the string tension $T = 1/(2\pi \alpha')$ where $l_s$ is the string length. For $N=1$ and to leading order in both expansions the low energy effective Lagrangian is  the abelian Born-Infeld Lagrangian \cite{Fradkin:1985qd,Abouelsaood:1986gd}%
\footnote{Note that since $\CF_{ab} = B_{ab} + 2\pi \alpha' F_{ab}$ one can regard the non-linear leading order Lagrangian \eqref{BI_LN1} as including an infinite number of $\alpha'$ corrections in terms of the worldvolume field strength $F_{ab}$. However, as we work in terms of $\CF_{ab}$ we shall think of all these terms as leading order in the $\alpha'$ expansion. }
\begin{equation}
\label{BI_LN1}
\CL = -  T_{\rm D3} \sqrt{- \det ( \eta_{ab} + \CF_{ab} ) }
\end{equation}
where $\eta_{ab} = \mbox{diag} (-1,1,1,1)$ is the four-dimensional Minkowski metric and the indices $a,b=0,1,2,3$ are world-volume indices. Moreover, $T_{\rm D3} = 1/(g_s l_s^4 (2\pi)^3 )$ is the D3-brane tension. 
Here $\CF_{ab} = B_{ab} + 2\pi l_s^2 F_{ab}$ is the gauge invariant field strength for the abelian Born-Infeld Lagrangian. We set the world-volume field strength $F_{ab}$ to zero. We do not have any Chern-Simons terms since the bulk field that would couple to these are zero in our setup. The field strength $\CF_{ab}$ obeys \eqref{BI_FS}.

Since we consider zero world-volume field strength the effective Lagrangian for any $N$ is 
\begin{equation}
\label{BI_LN}
\CL = -  N T_{\rm D3}  \sqrt{- \det ( \eta_{ab} + \CF_{ab} ) }
\end{equation}
which is just the abelian Born-Infeld Lagrangian \eqref{BI_LN1} multiplied by an overall factor of $N$. Thus, we are focussing on the abelian part of the non-abelian Born-Infeld Lagrangian. One can see this factor arising from the trace over the Chan-Paton factors for the open string. This trace is trivial as we set the non-abelian world-volume field strength to zero. We shall choose to be in the regime $N \gg 1$ for reasons mentioned below.

The Lagrangian \eqref{BI_LN} is computed from open strings with the world-sheet having the topology of a disc, 
\begin{figure}
\begin{center}
\includegraphics[scale=0.45]{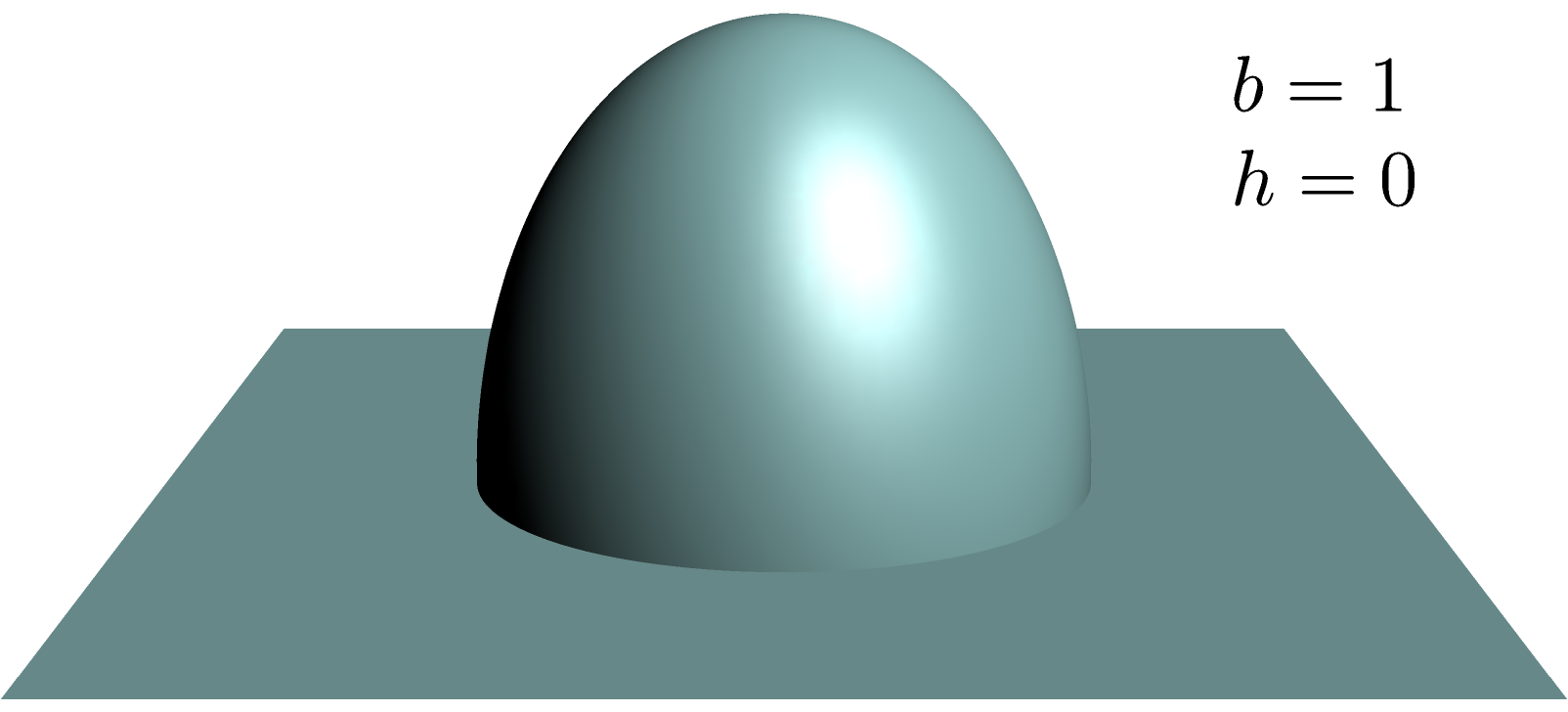}
\caption{Disc topology}
\label{fig:disk}
\end{center}
\end{figure}
see Figure \ref{fig:disk}. A slowly varying background Kalb-Ramond field introduces an interaction that induces higher-derivative corrections to the Lagrangian \eqref{BI_LN}.  Each derivative comes with a string length $l_s$. Since one can only write covariant terms with an even number of derivatives, one can see it as an expansion in $\alpha'$. We review these corrections in Section \ref{sec:BIside}.

\begin{figure}[ht]
\begin{center}
\includegraphics[scale=0.55]{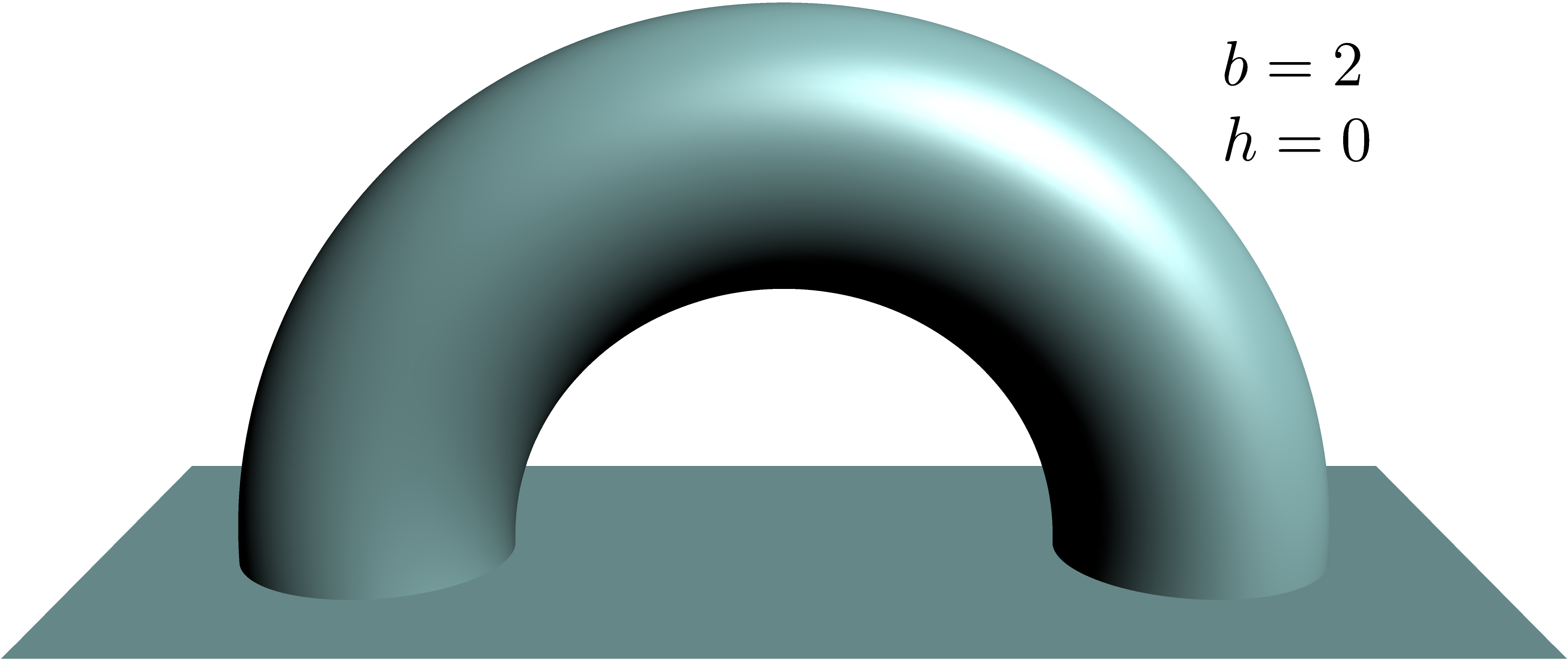}
\caption{World-sheet topology with two boundaries.}
\label{fig:twoboundaries}
\end{center}
\end{figure}

\begin{figure}[ht]
\begin{center}
\includegraphics[scale=0.6]{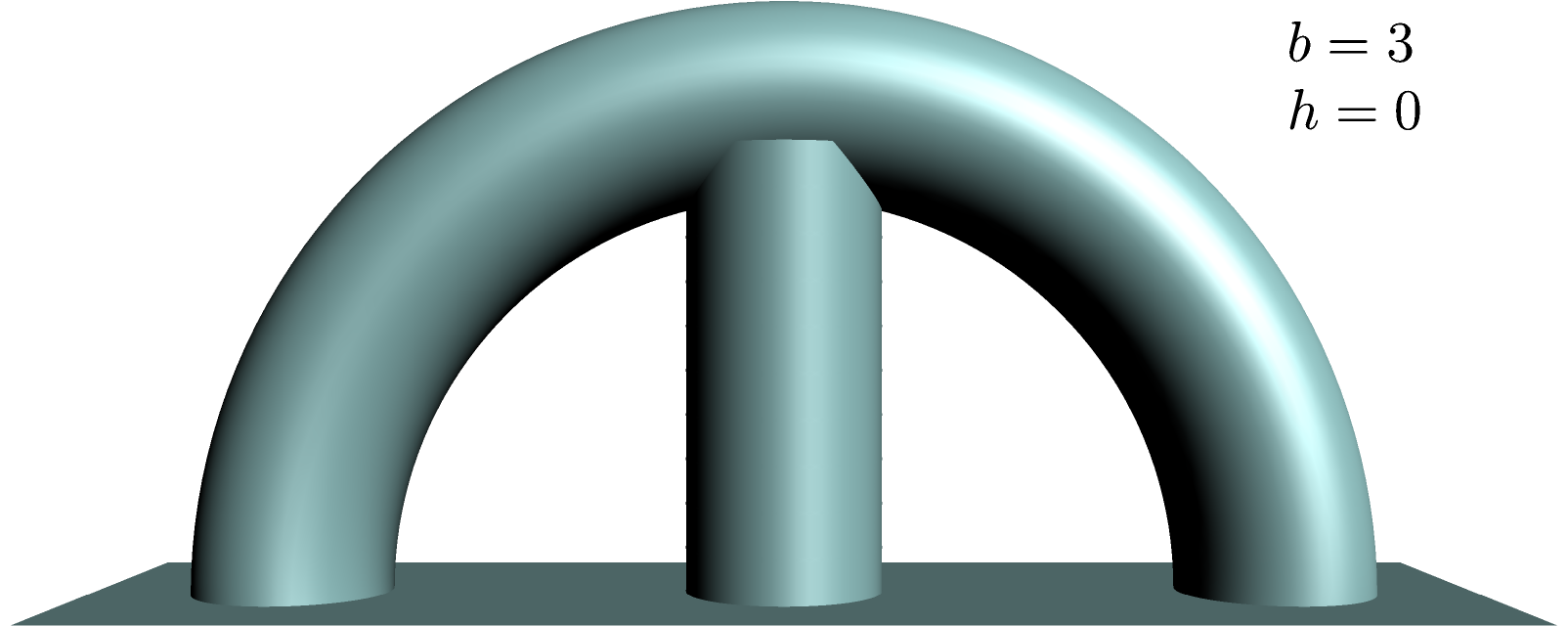}
\caption{World-sheet topology with three boundaries.}
\label{fig:threeboundaries}
\end{center}
\end{figure}

\begin{figure}[ht]
\begin{center}
\includegraphics[scale=0.6]{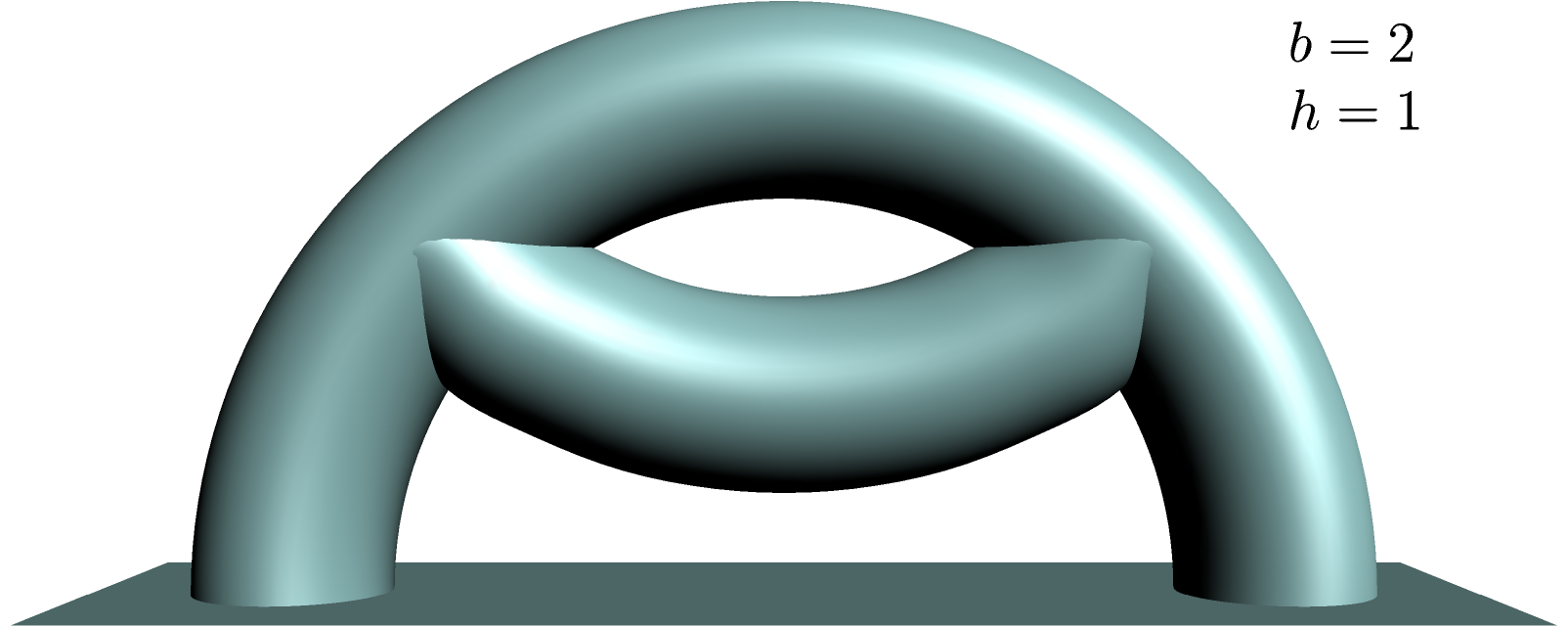}
\caption{World-sheet topology with thwo boundaries and one handle.}
\label{fig:onehandle}
\end{center}
\end{figure}

The other type of corrections to the Born-Infeld Lagrangian are from the topological expansion of the open string world-sheet. A given world-sheet topology comes with the following weight
\begin{equation}
\label{weight1}
g_s^{-\chi} N^b 
\end{equation}
where $\chi=2 - 2h - b$ is the Euler character, $h$ the number of handles and $b$ the number of boundaries of the world-sheet. The factor $N^b$ arises due to the open string Chan-Paton factors.%
\footnote{An alternative way to get the $N^b$ factor is to take the decoupling limit $\alpha' \rightarrow 0$. Then one has $U(N)$ Yang-Mills theory and the $N^b$ factor arises from $b$ loops with $N$ gluons.}
In particular, the disc topology has $h=0$ and $b=1$ thus giving the weight $g_s^{-1} N$. This matches the weight of the leading order Born-Infeld Lagrangian \eqref{BI_LN}.  In Figures \ref{fig:twoboundaries}, \ref{fig:threeboundaries} and \ref{fig:onehandle} we provide a few examples of world-sheet topologies. 

Since we would like to connect to the gravity description of the D3-branes we are interested in the corrections for large $N$ and weak string coupling $g_s \ll 1$ and thus $g_s N \gg g_s$. We can rewrite the weight as
\begin{equation}
\label{weight2}
\frac{N}{g_s} (g_s N)^{b-1} g_s^{2h}
\end{equation}
We see here that the leading corrections are the planar corrections with no handles $h=0$. The planar diagrams are thus obtained by adding more boundaries to the world-sheet without adding handles. As anticipated above in Section \ref{sec:setup}, we see that for our open string description to be valid we need 
\begin{equation}
g_s N \ll 1
\end{equation}
The planar diagrams with $h=0$ correspond to loop diagrams of the open string when $b > 1$. Indeed, the topology $h=0$ and $b=2$, illustrated in Figure \ref{fig:twoboundaries}, is a one-loop diagram for the open string. 

\subsubsection*{Closed string point of view}

Thanks to the open/closed duality of Polchinski \cite{Polchinski:1995mt}, one can equivalently compute the diagrams \ref{fig:twoboundaries}, \ref{fig:threeboundaries}, \ref{fig:onehandle} from a closed string point of view. For closed string theory, there is a D3-brane boundary state for each boundary on the string world-sheet \cite{DiVecchia:1999mal}. The closed string computation for each diagram gives the same result and the weight factor of Eqs.~\eqref{weight1}-\eqref{weight2} is the same where now $h$ is the number of handles on the closed string world-sheet and $b$ is the number of boundaries, each attached to a closed string boundary state. Thus, open/closed string duality enables one to compute the same diagram from both an open and a closed string point of view.

One can take into account the non-zero Kalb-Ramond field by finding the appropriate D3-brane boundary state \cite{DiVecchia:1999fje}. 

We now want to take the low energy limit and describe the physics using closed string variables, rather than open string variables. Hence we insert a vertex operator for any of the massless states of type IIB string theory on the diagrams. This enables us to describe the low energy limit of the diagram from the closed string point of view. 

\subsubsection*{Gravity description at strong coupling}

We now turn to the low energy effective description of this in terms of gravity. The low energy dynamics of closed strings is described by gravity plus matter. In this case type IIB supergravity. Inserting a single boundary state corresponds to considering the effect of a delta-function source for the D3-branes in a linearized gravity theory. Instead, as the coupling $g_s N$ becomes large, one should consider diagrams with an arbitrary number of boundaries $b$. In this way one approaches the fully backreacted solution where the gravity field is interacting with the source an arbitrary number of times. Indeed, the full non-linear gravity solution for $N$ D3-branes includes the sum over diagrams with an arbitrary large number of boundaries $b$. The insertion of the vertex operator for the massles closed string state corresponds to the fact that one is computing tree-level one point functions for the supergravity fields with an arbitrary number of interactions with the source.

Instead we only include the planar diagrams with $h=0$ as each handle, also from the closed string point of view, comes with a $g_s^2$ factor. To include such a correction one would need to go beyond type IIB supergravity. To suppress these corrections we need $g_s \ll 1$. 

As we shall review in Section \ref{sec:gravside}, the gravity description of the $N$ D3-branes is valid only for $g_s N \gg 1$, since we want to suppress the closed string $\alpha'$ corrections as well. Hence on the closed string side we are in the regime $g_s N \gg 1$ and $g_s \ll 1$. This we can regard as the strong coupling regime of the Born-Infeld action.

\subsubsection*{Holographic point of view}

Our above considerations implies two regimes in which we can find low energy effective descriptions of our D3-brane setup of Section \ref{sec:setup}. In the weakly coupled regime $g_s N \ll 1$ we can use the low energy effective description derived from open strings to obtain the effective action. Instead in the strongly coupled regime $g_s N \gg 1$ the low energy effective action can be obtained from gravity. In this sense we see that our correspondence is a holographic one, as we have a weak/strong coupling correspondence between a theory without gravity and one with gravity. From the open/closed string duality point of view, this merely corresponds to describing the same string diagrams in terms of a different set of variables.

\section{Born-Infeld side}
\label{sec:BIside}

As explained above, what we call the Born-Infeld side of the Born-Infeld/gravity correspondence is the effective low energy description of open string theory in the setup of Section \ref{sec:setup} with $N$ D3-branes in the background of a slowly varying Kalb-Ramond field. While we have $N \gg 1$ and $g_s \ll 1$ for both sides of the correspondence the Born-Infeld side gives a good description at weak coupling
\begin{equation}
g_s N \ll 1
\end{equation}
Below we describe the low energy effective action to zeroth order in $g_s N$. This includes the first few orders in the $\alpha'$ expansion, giving rise to higher derivative terms.

\subsubsection*{Leading order Born-Infeld action}

For our setup the leading order low energy effective action for $g_s N\ll 1$ and to leading order in the $\alpha'$ expansion is given by Eq.~\eqref{BI_LN}. We introduce now notation that will be useful for comparing with the strong coupling limit. Define 
\begin{equation}
\label{defineFGm}
\CF_a {}^b =  \CF_{ac}\eta^{cb} \spa G_{a} {}^b = \delta_a^b - \CF_a {}^c \CF_c {}^b \spa m = \sqrt{\det (\delta_a ^b + \CF_a {}^b)}  
\end{equation}
Below we shall regard $G_a {}^b$ and $\CF_a {}^b$ as 4 by 4 matrices. In particular, we define the inverse matrix $G^{-1}$ via $(G^{-1})_a {}^c G_c {}^b = \delta^b_a$. In this notation the Lagrangian \eqref{BI_LN} is $\CL = - N T_{\rm D3} m$. The equations of motion from this Lagrangian are
\begin{equation}
\label{BI_eom}
\partial^b ( m G^{-1} \CF)_{ba}  =0
\end{equation}
Here, as well as below, we use $\eta^{ab}$ and $\eta_{ab}$ to raise and lower indices.
The Born-Infeld energy-momentum tensor for the Lagrangian \eqref{BI_LN} is computed to be
\begin{equation}
\label{BI_EM2}
\tau_{ab} = -  N  T_{\rm D3} m (G^{-1})_{ab}
\end{equation}
Hence we can write the energy-momentum conservation as
\begin{equation}
\label{BI_EMcons}
\partial^b ( m (G^{-1})_{ba} ) = 0
\end{equation}

\subsubsection*{Higher derivative correction in $\alpha'$ expansion}

The $\alpha'$ expansion of the Born-Infeld action has been extensively studied, see for instance the excellent summary in \cite{Eenink:2005loa}. Here we shall consider the action only up to four derivatives, $i.e.$ to second order in $\alpha'$. 

Let us write the exact Lagrangian in a world-volume derivative expansion in the following fashion
\begin{equation}
\label{Lderexp}
\CL = \CL^{(0)} + \CL^{(2)} + \CL^{(4)} + \CO ( \partial^6 )\, ,
\end{equation}
where the zeroth order term, $\CL^{(0)}$, is the Born-Infeld Lagrangian \eqref{BI_LN}. As we
already mentioned only the even number of derivatives terms appear in the expansion. Each term $\CL^{(2n)}$ comes with a factor $\alpha' {}^n$.

The generic $2n$-derivative term in the expansion \eqref{Lderexp} can be further decomposed according to the number of $\CF$'s 
it contains
\begin{equation}
\label{LexpF}
\CL^{(2n)}=\sum_{m=2}^{\infty} \CL^{(2n,m)}\, ,
\end{equation}
where the index $m$ counts the number of $\CF$'s.

In ~\cite{Andreev:1988cb} it was shown that all the two-derivative terms are zero,
$\CL^{(2)} = 0$. 
It can be also proved that terms involving an odd number of $\CF$'s have to vanish  because of the invariance of 
the theory under worldsheet parity (see footnote in the Introduction of \cite{deRoo:2003xv}).
Furthermore all the $m=2$ terms in \eqref{LexpF} vanish as well~\cite{Eenink:2005loa}. So the sum in \eqref{LexpF} actually only runs
over the even $m$ starting from $m=4$. 

The four derivative term $\CL^{(4)}$ has been considered to all orders in $\CF$ \cite{Wyllard:2000qe} although an explicit form is not known. The contribution quartic in $\CF$ takes the form~\cite{Andreev:1988cb} 
\begin{equation}
\CL^{(4,4)} =  \frac{NT_{\rm D3}\, \alpha' {}^2 }{576} \left[ ( \eta^{ac} \eta^{bd} + 2 \eta^{ab} \eta^{cd}  ) M_{abcd} - \frac{1}{4} ( \eta^{ab} \eta^{cd} + 2 \eta^{ac} \eta^{bd}  ) K_{ab} K_{cd} \right]\, ,
\end{equation}
where $K_{ab}$ and $M_{abcd}$ are
\begin{equation}
K_{ab} = \partial_a \CF_{cd} \partial_b \CF^{cd} 
\spa 
M_{abcd} = \partial_a \CF_{e_1 e_2} \partial_b \CF^{e_2 e_3} \partial_c \CF_{e_3 e_4} \partial_d \CF^{e_4 e_1}\, .
\end{equation}
%

\section{Gravity side}
\label{sec:gravside}

Following Polchinski's conjecture \cite{Polchinski:1995mt} D-branes in the low energy limit of type IIB string theory source the Ramond-Ramond fields of type IIB supergravity. Hence as described in Section \ref{sec:OCduality} the setup of Section \ref{sec:setup} with $N$ D3-branes in the background of a slowly varying Kalb-Ramond field admits in this way a low energy effective theory description in terms of type IIB supergravity. This is a good description for
\begin{equation}
g_s N \gg 1
\end{equation}
Below we describe the aspects of type IIB supergravity that we need for the particular setup, which effectively becomes a five-dimensional gravitational theory. The leading order solution for arbitrary constant $\CF_{ab}$ is given in Section \ref{sec:zerosol} in a world-volume covariant form. In Section \ref{sec:duality_chain} we generate this solution from a chain of dualitites and we employ this in Section \ref{sec:uniqD3} to show uniqueness of the solution. In Section \ref{sec:corrected_sol} we consider the derivative expansion for a slowly varying $\CF_{ab}$ field and show that there are no two-derivative corrections to the corresponding strongly coupled Born-Infeld action. Finally in Section \ref{sec:reading_EM} we explain how the asymptotic region of the D3-branes is defined, given that one wants to have a derivative expansion, and how to read off the energy-momentum tensor of the D3-branes in the asymptotic region.

\subsection{Setup in ten and five dimensions}
\label{sec:5Dsetup}

The Lagrangian of the bosonic part of type IIB supergravity is%
\footnote{The Chern-Simons term equals $\frac{1}{2} ( A^{(4)} \wedge H^{(3)} \wedge F^{(3)} )_{01\cdots 9}$.}
\begin{eqnarray}
\label{LagrtypeIIB}
\CL &=& \sqrt{-g} \Big( \CR^{(10)} - \frac{1}{2} \partial_\mu \phi \partial^\mu \phi - \frac{1}{12} e^{-\phi} H_{\mu\nu\rho} H^{\mu\nu\rho} - \frac{1}{2} e^{2\phi} \partial_\mu \chi \partial^\mu \chi - \frac{1}{12} e^\phi F_{\mu\nu\rho} F^{\mu\nu\rho} \nn \\[3mm] && - \frac{1}{4 \cdot 5!} F_{\mu\nu\rho\lambda \sigma}F^{\mu\nu\rho\lambda \sigma} \Big)
+ \frac{1}{8 \cdot 4!} e^{\mu_1 \cdots \mu_{10}} A_{\mu_1 \cdots \mu_4} \partial_{\mu_5} B_{\mu_6\mu_7} \partial_{\mu_8} A_{\mu_9 \mu_{10}}
\end{eqnarray}
Here $\phi$ is the dilaton and $\chi$ is the axion (or zero-form RR potential). The field strengths are defined as $H^{(3)} = dB^{(2)}$, $F^{(3)} =  dA^{(2)} - \chi \wedge H ^{(3)}$ and $F^{(5)} =  dA^{(4)} - A^{(2)} \wedge H ^{(3)}$ where $A^{(2)}$ and $A^{(4)}$ are the RR two- and four-form potentials, respectively, while $B^{(2)}$ is the Kalb-Ramond two-form potential. The metric  $g_{\mu\nu}$ is in the Einstein frame and $\CR^{(10)}$ is the corresponding Ricci scalar. The Hodge dual of the five-form field strength is defined as $\hod F_{\mu_1 \cdots \mu_{5}} = \frac{1}{5!} \sqrt{-g} \epsilon_{\mu_1 \cdots \mu_{10}} F^{\mu_6 \cdots \mu_{10}}$. 
The self-duality constraint of type IIB supergravity is $\hod F^{(5)} = F^{(5)}$. 

Consider now the supergravity description of $N$ D3-branes in the background of a slowly varying Kalb-Ramond field, as described above. $r$ denotes here the radial coordinate in the transverse space of the D3-branes, hence they are situated at $r=0$. The boundary condition of having a slowly varying Kalb-Ramond field as background can be formulated as
\begin{equation}
\label{KRfield}
 \lim_{r\rightarrow 0} B_{ab}= 0  \spa \lim_{r\rightarrow \infty} B_{ab} = \CF_{ab}
\end{equation}
with $\CF_{ab}(x^c)$ obeying \eqref{BI_FS}.
The metric asymptotes to ten-dimensional Minkowski space while the scalars $\phi$, $\chi$ and the RR two-form potential $A_{(2)}$ asymptote to zero for $r \rightarrow \infty$. As the D3-branes are situated at $r=0$ the ten-dimensional metric takes the form
\begin{equation}
\label{10D_ansatz1}
g_{\mu\nu} dx^\mu dx^\nu = e^{\frac{1}{2}\eta} \left( h_{mn} dx^m dx^n +  r^2 d\Omega_5^2 \right) 
\end{equation}
where $m,n=0,1,2,3,4$ are indices for the five-dimensional metric $h_{mn}$ and $\eta$ is a scalar. Note that $h_{mn}$ and $\eta$ only depends on $x^m$, $m=0,...,4$.This follows from using the symmetries of the setup as well as the gauge freedom of the ten-dimensional metric. One can furthermore impose the following gauge for the five-dimensional metric
\begin{equation}
\label{10D_ansatz2}
h_{mn} dx^m dx^n = h_{ab} dx^a dx^b + dr^2
\end{equation}
where $r = x^4$ and $a,b=0,1,2,3$ denotes the indices for the world-volume directions. To impose that we have $N$ D3-branes we set the five-form RR field strength to be
\begin{equation}
\label{F5ans}
F^{(5)} =  4 r_c^4 ( \omega_5 +\hod \omega_5  )
\end{equation}
where $\omega_5$ is the volume-form for the unit five-sphere and $r_c$ is defined as
\begin{equation}
\label{charge_radius}
r_c^4 = \frac{N}{2\pi^2 T_{\rm D3}}  
\end{equation}
As we shall see below $r_c$ sets the length scale of the thickness of the D3-brane.

Since only the metric and $F_{(5)}$ involve the five-sphere directions we can effectively reduce the above ten-dimensional setup to a five-dimensional one. To this end, write the ten-dimensional metric in slightly more general form
\begin{equation}
\label{10metBans}
g_{\mu\nu} dx^\mu dx^\nu = e^{\frac{1}{2}\eta} \left( h_{mn} dx^m dx^n +  e^{2\rho} d\Omega_5^2 \right) 
\end{equation}
introducing the additional scalar field $\rho$.
Taking this together with the ansatz \eqref{F5ans} for $F_{(5)}$ we can instead find the gravitational background that corresponds to the D3-branes as a solution of a five-dimensional theory of gravity with metric $h_{mn}$ plus matter fields $\phi$, $\chi$, $\eta$, $\rho$, $B_{mn}$ and $A_{mn}$. In this five-dimensional reduced theory with coordinates $x^m$ and metric $h_{mn}$ the EOMs for the matter fields are
\begin{equation}
\label{5Deq1}
\begin{array}{c} \ds
\frac{\partial_m ( \sqrt{-h} \, e^{2\phi + 2 \eta + 5\rho} \, \partial^m \chi  )}{e^{5\rho} \sqrt{-h}} = - \frac{1}{6} e^{\phi + \eta} F \cdot H
\\[5mm] \ds
 \frac{\partial_m (  \sqrt{-h} \, e^{2\eta+5\rho} \, \partial^m \phi )}{e^{5\rho}\sqrt{-h}}   = - \frac{1}{12} e^{\eta-\phi} H^2 + \frac{1}{12} e^{\eta+\phi} F^2 + e^{2\eta+2\phi} (\partial \chi )^2
\\[5mm] \ds
\frac{4}{e^{2\rho}} -5 (\partial \rho)^2 -\frac{1}{2} (\partial \eta)^2 -\frac{13}{4} \partial \eta \cdot \partial \rho - D^2 \rho- \frac{1}{4} D^2  \eta  = - \frac{e^{-\eta}}{48} (e^{-\phi} H^2 + e^{\phi} F^2) + \frac{4 r_c^8}{e^{2\eta+10\rho}} 
\\[4mm] \ds
 \partial_m \Big( \sqrt{-h} \, e^{\eta + 5\rho} (e^{-\phi} H^{mnk} - \chi e^\phi F^{mnk} )\Big) = - \frac{4 r_c^4}{6} \epsilon^{nklpq} (F_{lpq} + \chi H_{lpq})
\\[4mm] \ds
\partial_m \Big( \sqrt{-h} \, e^{\eta + 5\rho}  e^{\phi} F^{mnk} \Big) = \frac{4 r_c^4}{6} \epsilon^{nklpq}  H_{lpq}
\end{array}
\end{equation} 
where $F^2= F_{mnk}F^{mnk}$,  $F \cdot H = F_{mnk} H^{mnk}$, $(\partial \chi)^2 = \partial_m \chi \partial^m \chi$, $\partial \eta \cdot \partial \rho = \partial_m \eta \partial^m \rho$ and $D^2 \eta = D^m D_m \eta$. The Einstein equations take the form
\begin{eqnarray}
\label{5Deq2}
&& \CR_{mn} =  5 \partial_m \rho \partial_n \rho + 5  D_m D_n \rho  - \frac{1}{2}\partial_m \eta \partial_n \eta + 2 D_m D_n \eta + \frac{1}{2} h_{mn} (\partial \eta)^2 \nn \\ && + \frac{1}{4} h_{mn}  D^2 \eta + \frac{5}{4}  h_{mn}   \partial \eta \cdot \partial \rho + \frac{1}{2} \partial_m \phi \partial_n \phi + \frac{1}{2} e^{2\phi} \partial_m\chi \partial_n \chi -  4 r_c^8 e^{-2\eta - 10 \rho}  h_{mn}  \nn \\ && +  \frac{1}{12} e^{-\phi-\eta} (3 H_m {}^{kl}H_{nkl} - \frac{1}{4}  h_{mn}   H^2)  + \frac{1}{12} e^{\phi-\eta} (3 F_m {}^{kl}F_{nkl} - \frac{1}{4}  h_{mn}   F^2 )  
\end{eqnarray}
where $\CR_{mn}$ is the Ricci tensor and $D_m$ is the covariant derivative with respect to $h_{mn}$.

\subsection{General solution for D3-branes with constant $\CF_{ab}$ field}
\label{sec:zerosol}

We now present the general solution for $N$ D3-branes with the boundary condition \eqref{KRfield} on the Kalb-Ramond two-form where $\CF_{ab}$ is constant. This is a new solution. It generalizes the previously known solutions of the $D3 \parallel (F1 \parallel D1)$ bound state \cite{Cederwall:1998tr,Lu:1999uv} corresponding to $e.g.$ $\CF_{01}$ and $\CF_{23}$ turned on, and $D3 \parallel (F1 \perp D1)$ bound state \cite{Grignani:2013ewa} corresponding to $e.g.$ $\CF_{03}$ and $\CF_{23}$ turned on.%
\footnote{We note that for the case in which the D3-brane world-volume is on four-dimensional Minkowski-space, the classification of the types of possible bound states is equivalent to the classification of constant electromagnetic field strength configurations, since $\CF_{ab}$ is the same type of field. From this one learns that using boosts and rotations one can reach either the case in which the only non-zero components are $\CF_{01}$ and $\CF_{23}$, including the cases in which one of these are zero, or the case $\CF_{01}=\pm \CF_{12}$, corresponding to having an orthogonal electric and magnetic field equal in magnitude. Thus, using boosts and rotations, all configurations can be reached starting from either the $D3 \parallel (F1 \parallel D1)$ or the $D3 \parallel (F1 \perp D1)$ bound states. Of course, in our case it is crucial to have constructed the explicit solution for arbitrary $\CF_{ab}$ since we allow in the following $\CF_{ab}$ to vary over the world-volume. Another instance in which it is crucial to have our explicit solution is when compactifying on $T^3$.}

We use in the following the four-dimensional Minkowski metric $\eta_{ab}$ as well as $\CF_a {}^b$, $G_a {}^b$ and $m$ defined in \eqref{defineFGm}. We regard $\CF_a {}^b$ and $G_a {}^b$ as 4 by 4 matrices. We define furthermore
\begin{equation}
\hod \CF_{ab} = \frac{1}{2} \epsilon_{abcd} \CF^{cd} 
\end{equation}
Note that indices on $G_a {}^b$ and $\CF_a {}^b$, $\hod \CF_a {}^b$ and their matrix products in the following always are raised and lowered with $\eta^{ab}$ and $\eta_{ab}$. One can compute that $\tr (G)  = 4 - \tr (\CF^2)$ and $\det (G) = m^4$. The ansatz \eqref{10D_ansatz1}-\eqref{10D_ansatz2}  for the ten-dimensional metric is translated into 
\begin{equation}
\label{5D_ansatz}
h_{mn} dx^m dx^n =   h_{ab}  dx^a dx^b+ dr^2  \spa \rho = \log r
\end{equation}
The solution is 
\begin{equation}
\label{Sol0}
\begin{array}{c}\ds
h_{ab} = \{  ( I + m r_c^4 r^{-4} G^{-1} )^{-1} \}_a {}^c \eta_{cb}  \spa \sqrt{-h} = e^{-2\eta}
\\[4mm]\ds
e^{2\eta} =  1 + \frac{\tr (G)}{2m}  \frac{r_c^4}{r^4}+ \frac{r_c^8}{r^8}
\spa
e^{2\phi} = e^{-2\eta} \Big(1 + m\frac{r_c^4}{r^4}\Big)^2
\spa
\chi =\frac{\tr ( \CF \, \hod \CF )}{4}  \frac{\frac{r_c^4}{r^4}}{1 + m \frac{r_c^4}{r^4}}
\\[4mm]\ds
B_{ab} =  \CF_a {}^c h_{cb}
\spa 
A_{ab} = \frac{r_c^4}{r^4} \, \hod \CF_{a} {}^c h_{cb}  
\end{array}
\end{equation}
where $r_c$ is given in \eqref{charge_radius}.

The above is the solution for $N$ D3-branes with a constant $\CF_{ab}$. Since this should provide a low energy description of the D3-brane dynamics we should have $r_c \gg l_s$. This requires
\begin{equation}
g_s N \gg 1
\end{equation}
which reflects that the above solution is the low energy effective description of the strongly coupled Born-Infeld theory. 

\subsection{Duality chain generating constant $\CF_{ab}$ D3-brane solution}
\label{sec:duality_chain}

The solution \eqref{Sol0} for $N$ D3-branes with a constant $\CF_{ab}$ field can be generated by a chain of T-dualities, boosts and rotations. This reveals in which sense it generalizes the $D3 \parallel (F1 \parallel D1)$ bound state \cite{Lu:1999uv} and the $D3 \parallel (F1 \perp D1)$ bound state \cite{Grignani:2013ewa} solutions. 

Begin with the D3-brane solution with $\CF_{ab}=0$
\begin{equation}
\label{D3brane_sol}
h_{ab}= \frac{\eta_{ab}}{1 + \frac{r_c^4}{r^4}} \spa e^{\eta} = 1 + \frac{r_c^4}{r^4} \spa \phi = \chi = 0 \spa B_{ab}=A_{ab}=0
\end{equation}
T-dualize along $x^1$, giving D2-branes smeared along $x^1$. Rotate this in the $12$-plane and T-dualize again along $x^1$ revealing a D1-D3 bound state with the D-strings along the $x^3$-direction. T-dualize this along $x^3$, giving a D0-D2 bound state smeared along $x^3$-direction. Rotate this in the $23$-plane, boost along the $x^3$-direction and finally T-dualize along the $x^3$-direction. This reveals a bound-state of D3-branes, F-strings in the $x^3$-direction and D-strings along a direction in the $13$-plane. In this way one finds the solution \eqref{Sol0} with $\CF_{03}\neq 0$, $\CF_{12}\neq 0$ and $\CF_{23}\neq 0$. One can now proceed using rotations in the $12$-plane, $13$-plane and $23$-plane to find the solution \eqref{Sol0} with general constant $\CF_{ab}$. Note that the solution that arises from the above transformations has
\begin{equation}
\label{KRfield2}
 \lim_{r\rightarrow 0} B_{ab}= - \CF_{ab}  \spa \lim_{r\rightarrow \infty} B_{ab} =0
\end{equation}
Hence, one should make the appropriate gauge transformation of the Kalb-Ramond field to arrive at the one  written in \eqref{Sol0}. 

\subsection{Uniqueness of constant $\CF_{ab}$ D3-brane solution}
\label{sec:uniqD3}

For use below we prove in this subsection the perturbative uniqueness of the constant $\CF_{ab}$ D3-brane solution \eqref{Sol0}. Clearly we can demand that the solution \eqref{Sol0} plus any possible perturbation should be translationally invariant along the four world-volume directions, as well as rotationally symmetric in the transverse directions.

Consider first the perturbative uniqueness of the D3-brane solution with $\CF_{ab}=0$, $i.e.$ in a background without any field or potential turned on. This solution is \eqref{D3brane_sol}.
Consider now perturbing the D3-brane solution by a small fluctuation. We demand that the fluctuation is translationally symmetric along the world-volume directions $x^a$, $a=0,1,2,3$, as well as rotationally symmetric in the orthogonal directions to the D3-brane. This means we can describe the fluctuation using the five-dimensional setup of Sec.~\ref{sec:5Dsetup} and in particular the metric gauge \eqref{5D_ansatz}. Moreover, we can work in a gauge where the fluctuation only depends on the $r$ coordinate. Thus to any field or component of field $X_I=(h_{ab},\eta,\phi,\chi,B_{ab},A_{ab})$ we add a fluctuation $\delta X_I (r)=( \delta h_{ab} (r),\delta \eta (r),\delta \phi (r),\delta \chi (r),\delta B_{ab}(r),\delta A_{ab}(r))$. We now put this into the equations of motion \eqref{5Deq1}-\eqref{5Deq2}. Since the fluctuation is assumed to be small we can linearize the equations for the fluctuations. After some straightforward algebra one finds that the most general fluctuation is of the form
\begin{equation}
\label{D3brane_fluc}
\begin{array}{c}\ds
\delta \phi = \frac{A_\phi}{r^4} + B_{\phi} \spa \delta \chi = \frac{A_\chi}{r^4} + B_{\chi} \spa \delta \eta = \frac{B_\eta r^{12} + A_{\eta} ( 5r^8 + 4r_c^4 r^8 + r_c^8 )}{r^8 (r^4 + r_c^4)}
\\[4mm]\ds
\delta B_{ab} = \frac{U^{(1)}_{ab}}{r^4} +\frac{U^{(2)}_{ab}}{r^4+r_c^4} + U^{(3)}_{ab} \spa \epsilon^{abcd} \delta A_{cd} = - \frac{U^{(1)}_{ab}}{r^4} +\frac{U^{(2)}_{ab}}{r^4+r_c^4} + U^{(4)}_{ab}
\\[4mm]\ds
\delta h_{ab} =  -r_c^4 \frac{ 2 B_\eta r_c^4 r^4 + A_\eta (3 r_c^4 - r^4)}{2 r^4
   \left(r_c^4+r^4\right)^2} \eta_{ab}+ \frac{V^{(1)}_{ab}}{r^4 + r_c^4} + V^{(2)}_{ab} 
\end{array}
\end{equation}
Here we introduced several integration constants $A_i$, $B_i$, $i=\phi,\chi,\eta$, $U^{(j)}_{ab}$, $j=1,2,3,4$ and $V^{(k)}_{ab}$, $k=1,2$, where $a,b=0,1,2,3$, $U^{(j)}_{ab}$ are antisymmetric and $V^{(k)}_{ab}$ are symmetric.

Demand now that the solution plus fluctuation should be asymptotically flat for $r \rightarrow \infty$. This gives immediately $B_{\phi}=B_{\chi}=B_{\eta}=0$ and $V^{(2)}_{ab}=0$. Since we also demand that $B_{ab}$ and $A_{ab}$ go to zero for $r\rightarrow \infty$ we get in addition $U^{(3)}_{ab} = U^{(4)}_{ab}=0$.

Consider instead what happens for $r\rightarrow 0$. Our fluctuation is assumed to be small. However, if we for instance consider the dilaton of \eqref{D3brane_fluc} we see that if $A_\phi\neq 0$ then the fluctuation will blow up in comparison to the solution \eqref{D3brane_sol} as $r \rightarrow 0$, even if one had a constant non-zero value of $\phi$. Thus, one can conclude $A_\phi=0$. From the same line of arguments one finds $A_\chi=A_\eta=0$, $V^{(1)}_{ab}=0$ and $U^{(1)}_{ab}=U^{(2)}_{ab}=0$. Hence the fluctuation is zero and we have shown perturbative uniqueness of the D3-brane solution.  

We can now use the duality chain of Sec.~\ref{sec:duality_chain} to show uniqueness in the general case. Start with the D3-brane solution \eqref{Sol0} with an arbitrary constant $\CF_{ab}$. We write this solution as $Y^{(0)}_I=(h_{ab},\eta,\phi,\chi,B_{ab},A_{ab})$ where $I$ runs over all the fields and their components. Consider adding a fluctuation $\delta Y_I (r)=( \delta h_{ab} (r),\delta \eta (r),\delta \phi (r),\delta \chi (r),\delta B_{ab}(r),\delta A_{ab}(r))$  to this solution with translational symmetry along the world-volume and rotational symmetry in the transverse directions. For use below we record that the fluctuations obey a set of equations of the form 
\begin{equation}
\label{fluc_eq}
H^{m,I}_{(0)} \delta Y_I + H^{m,I}_{(1)} \partial_r \delta Y_I + H^{m,I}_{(2)} \partial_r^2 \delta Y_I = 0
\end{equation}
where $m$ runs over all the equations of motion \eqref{5Deq1}-\eqref{5Deq2} and one sums $I$ and $J$ over all the fields and their components.
We require that the solution plus fluctuation $Y_I = Y_I^{(0)} + \delta Y_I$ is asymptotically flat and that the fluctuation is small everywhere. Make a gauge transformation of the Kalb-Ramond field such that the boundary condition \eqref{KRfield} instead is \eqref{KRfield2}. Follow now in reverse order the chain of rotations, T-dualities and boosts of Sec.~\ref{sec:duality_chain}. After these transformations we end up with the D3-brane solution \eqref{D3brane_sol} (one should transform the Kalb-Ramond field to the appropriate gauge) obtained by acting with the transformations on \eqref{Sol0} along with a transformed fluctuation which must be of the general form \eqref{D3brane_fluc}. One can easily see that demanding asymptotic flatness before and after the transformations is equivalent. Hence the fluctuation must be zero by our above arguments.  Thus, this shows perturbative uniqueness of the D3-brane solution \eqref{Sol0} with an arbitrary constant $\CF_{ab}$ field.

\subsection{Corrected solution for slowly varying $\CF_{ab}$ field}
\label{sec:corrected_sol}

We would like to describe the strong coupling limit of Born-Infeld theory with a varying $\CF_{ab}$ field along the world-volume. Above we saw that using the gravitational description as a low energy effective theory requires $r_c  \gg l_s$. Since we want to approach finding the effective action using an expansion in world-volume derivatives of $\CF_{ab}$ we need that $\CF_{ab}$ varies sufficiently slowly. Since the thickness of the D3-branes is of order $r_c$ a perturbative expansion in world-volume derivatives requires that $R \gg r_c$ since we can then use the constant $\CF_{ab}$ solution \eqref{Sol0} as a valid solution for length scales smaller than of order $R$. 

Considering the equations of motion \eqref{5Deq1}-\eqref{5Deq2} we now include the world-volume derivatives of the fields. The ansatz \eqref{5D_ansatz} is used at any order. The solution \eqref{Sol0} then corresponds to the zeroth order solution that we expand around. 

It is important to note that the perturbatively corrected solution that we are considering below only can be valid for $r \ll R$. This is because an observer at $r \gg R$ would get contributions from a piece of the brane which is larger than the minimal scale of the variation of $\CF_{ab}$. Hence for this observer the corrections to \eqref{Sol0} would be of the same order as the solution \eqref{Sol0}.

\subsubsection*{First order in derivative expansion}

We start by looking at the equations that involve a single world-volume derivative. On general grounds we expect that the solution \eqref{Sol0} cannot be corrected by terms with an odd number of world-volume derivatives. This is because all the fields involved either have zero ($\eta$, $\phi$ and $\chi$) or two ($h_{ab}$, $B_{ab}$ and $A_{ab}$) world-volume indices. Since $\CF_{ab}$ has two world-volume indices it is not possible to write down terms that are special-relativistically covariant on the world-volume and that have an odd number of world-volume derivatives. However, the equations are still non-trivial as we now shall see.

Eqs.~\eqref{5Deq1} contains two equations with one world-volume derivative
\begin{equation}
\label{coneq1}
\begin{array}{c}\ds
\partial_b \Big(  e^{-\eta-\phi} H^{bar} - \chi e^{-\eta+\phi} F^{bar} \Big) =  \frac{4 r_c^4}{6r^5} \epsilon^{abcd} (F_{bcd} + \chi H_{bcd})
\\[4mm]\ds
\partial_b \Big(  e^{-\eta+\phi} F^{bar} \Big) = - \frac{4 r_c^4}{6 r^5} \epsilon^{abcd}  H_{bcd}
\end{array}
\end{equation} 
while Eq.~\eqref{5Deq2} contains one equation
\begin{eqnarray} 
\label{coneq2}
 \frac{1}{2} \partial_b ( h^{bc} \partial_r h_{ca} )   - \frac{1}{2}\Gamma^c_{ab} h^{bd}\partial_r h_{dc} &=&   - \frac{1}{2}\partial_a \eta \partial_r \eta     + \frac{1}{2} \partial_a \phi \partial_r \phi + \frac{1}{2} e^{2\phi} \partial_a\chi \partial_r \chi \nn \\ &&  +  \frac{1}{4} e^{-\phi-\eta}  H_a {}^{bc}H_{bcr}   + \frac{1}{4} e^{\phi-\eta}  F_a {}^{bc}F_{bcr}  
\end{eqnarray}
here written imposing $\log \sqrt{-h}=-2\eta$.

We first consider the above equations to leading order at large $r$. For the first equation of Eqs.~\eqref{coneq1} we see that the LHS starts at order $1/r^5$ while the RHS at order $1/r^9$. Hence to leading order we have to satisfy $\partial_b H^{bar}=0$. This gives the constraint
\begin{equation}
\label{conKR}
\partial^b ( m G^{-1} \CF)_{ba} = 0
\end{equation}
Using the same reasoning for the second equation of Eqs.~\eqref{coneq1} one finds to leading order at large $r$ that $\partial_b F^{bar}=0$. This gives $\partial^b ( \hod \CF)_{ba} = 0$ which is equivalent to $d\CF=0$.
For Eq.~\eqref{coneq2} we find to leading order at large $r$ that $\partial^b \partial_r h_{ba} = 0$. This gives the constraint
\begin{equation}
\label{conEE}
\partial^b ( m G^{-1} )_{ba} = 0
\end{equation}
It is straightforward to verify explicitly that the three equations of Eqs.~\eqref{coneq1}-\eqref{coneq2} for any $r$ are obeyed assuming we impose \eqref{conKR} and \eqref{conEE} as constraints on the variation of $\CF_{ab}$.

Thus, we have derived from the five-dimensional gravity theory that allowing $\CF_{ab}$ to vary slowly along the world-volume leads to the constraints \eqref{conKR}-\eqref{conEE} on the variation of $\CF_{ab}$. This matches perfectly with the weakly coupled Born-Infeld theory description, since we recognise \eqref{conKR} as the Born-Infeld equation of motion \eqref{BI_eom} and \eqref{conEE} as the Born-Infeld energy-momentum conservation \eqref{BI_EMcons}.

\subsubsection*{Second order in derivative expansion}

At second order in the world-volume derivative expansion one has corrections to the zeroth order solution \eqref{Sol0}. Write the general solution including higher-derivative correction as $Y_I=(h_{ab},\eta,\phi,\chi,B_{ab},A_{ab})$ where the index $I$ runs over all the possible fields and their components of the solution. Then $Y_I$ is expanded in powers of the world-volume derivatives
\begin{equation}
Y_I = Y_I^{(0)} + Y_I^{(2)} + Y_I^{(4)} + \CO (\partial^6)
\end{equation}
where $Y_I^{(0)}$ refers to the zeroth order solution \eqref{Sol0}, $Y_I^{(2)}$ to the 2nd order correction with two world-volume derivatives acting on functions of $\CF_{ab}$ and $Y_I^{(4)}$ is similarly the 4th order correction, and so on. There are no corrections with an odd number of world-volume derivatives since one cannot write a covariant expression for this in terms of $\CF_{ab}$. The equations that determine $Y_I^{(2)}$ are all the equations in \eqref{5Deq1}-\eqref{5Deq2} involving an even number of world-volume derivatives. Schematically, one can write them as
\begin{equation}
\label{inhomoeq}
H^{m,I}_{(0)} Y^{(2)}_I + H^{m,I}_{(1)} \partial_r Y^{(2)}_I + H^{m,I}_{(2)} \partial_r^2 Y^{(2)}_I = K^{m,ab,I} \partial_a \partial_b Y^{(0)}_I + L^{m,ab,IJ} \partial_a Y^{(0)}_I \partial_b Y^{(0)}_J
\end{equation}
where $m$ runs over all the equations of motion and one sums $I$ and $J$ over all the fields and their components.

It is straightforward to find a solution to \eqref{inhomoeq} by working in a large $r$ expansion of $Y^{(2)}_I$. This reveals a solution of the form
\begin{equation}
\label{inhomosol}
Y^{(2)}_I = \sum_{n=0}^\infty \frac{C_{I,n}}{r^{2+4n}}
\end{equation}
Here the leading contributions are 
\begin{align}
			\chi^{(2)} &= \frac{r_c^4}{16 r^2} \partial_a \partial^a \tr ( \CF \, \hod \CF)+ \CO (r^{-6} ) &
			 \phi^{(2)} &= \frac{r_c^4}{4 r^2} \partial_a \partial^a \Big( m - \frac{\tr G}{4m} \Big)+ \CO (r^{-6} )\nn \\ 
			 B_{ab}^{(2)} &= - \frac{r_c^4}{4 r^2} 3 \partial^c \partial_{[c} ( m G^{-1} \CF )_{ab]} + \CO (r^{-6} )&
			 A_{ab}^{(2)} &=  \frac{r_c^4}{4 r^2} 3 \partial^c \partial_{[c} \hod \CF _{ab]} + \CO ( r^{-6} )\nn \\
			\eta^{(2)}& =  \frac{r_c^4}{16r^2} \partial^c \partial_c \Big( \frac{\tr G}{ m} \Big) + \CO ( r^{-6} ) &
			h_{ab}^{(2)} &= -  \frac{r_c^4}{4r^2} \partial^c \partial_c ( m G^{-1} )_{ab} + \CO ( r^{-6} )
\label{inhomosol_leading}
\end{align}

Given these leading contributions at order $1/r^2$ one can straightforwardly compute the $1/r^6$ contributions, and so forth, computing the full series \eqref{inhomosol}. Thus, this gives a particular solution to the \eqref{inhomoeq}. One can also find other solutions to \eqref{inhomoeq}, namely by adding a homogenous solution $\delta Y_I$ to the particular solution \eqref{inhomosol}. A homogenous solution $\delta Y_I$ solves \eqref{fluc_eq}. However, following the line of arguments of Sec.~\eqref{sec:uniqD3}, the general form for $\delta Y_I$ can be generated from \eqref{D3brane_fluc} by following the chain of dualities given in Sec.~\eqref{sec:uniqD3}. This shows that if we demand $\delta Y_I$ to be a small perturbation of $Y_I^{(0)}$ and that the whole solution should be asymptotically flat, then $\delta Y_I =0$. Hence \eqref{inhomosol} with leading $1/r^2$ terms \eqref{inhomosol_leading} is the full two-derivative contribution to the zeroth order solution \eqref{Sol0}. Below in Sec.~\eqref{sec:reading_EM} we use this to show that the strongly coupled Born-Infeld action does not have any derivative correction at the two-derivative order.

\subsubsection*{Third order in derivative expansion}
From the first of  \eqref{coneq1} we see that the LHS for the two-derivative corrections starts at order $1/r^2$ while the RHS at order $1/r^6$. Hence to leading order we have to satisfy $\partial_b H^{bar}=0$. This gives the constraint
\begin{equation}
\partial_b \partial^c \partial_{[c} ( m G^{-1} \CF )_{ab]} = 0
\end{equation}
which is satisfied assuming the constraint \eqref{conKR} is.

At leading order in the large $r$ expansion from  the second of \eqref{coneq1} we see that the LHS for the two-derivative corrections starts at order $1/r^2$ while the RHS at order $1/r^6$. Hence to leading order we have to satisfy $\partial_b F^{bar}=0$. This gives the constraint
\begin{equation}
\partial_b \partial^c \partial_{[c} \hod \CF _{ab]}  = 0
\end{equation}
which is satisfied assuming the constraint \eqref{BI_FS} is.

	From \eqref{coneq2} at leading order in large $r$ we get from the LHS $\partial^b \partial_r h_{ba} = 0$, this gives the constraint
\begin{equation} 
\partial^b \partial^c\partial_c( m G^{-1} )_{ba} = 0 
\end{equation}
which is of course satisfied assuming \eqref{conEE}.

\subsubsection*{Fourth order in derivative expansion}

The fourth order corrections at large $r$ can be put in the form
		\begin{equation}
		\label{YI4}
			Y_I^{(4)} = C_{I,0}^{(4)} \log r + \frac{C_{I,4}^{(4)}}{r^4}+ \CO\left(\frac{1}{r^8}\right)
		\end{equation}
One can easily compute  the leading order coefficients $C_{I,0}^{(4)}$, which read
			\begin{align}
			\label{YI42}
				C_{\chi,0}^{(4)} &= -\frac{r_c^4}{64} \partial^2 \partial^2 \tr  ( \CF \, \hod \CF) &
				C_{\phi,0}^{(4)} &= -\frac{r_c^4}{16} \partial^2 \partial^2 \Big( m - \frac{\tr G}{4m} \Big) \nn \\
				C_{B_{ab},0}^{(4)} &= \frac{r_c^4}{16} 3 \partial^2\partial^c \partial_{[c} ( m G^{-1} \CF )_{ab]} &
				C_{A_{ab},0}^{(4)} &= -\frac{r_c^4}{16} 3 \partial^2\partial^c \partial_{[c} \hod \CF _{ab]} \nn \\
				C_{\eta,0}^{(4)}& =  -\frac{r_c^4}{64} \partial^2 \partial^2 \Big( \frac{\tr G}{ m} \Big)  &
				C_{h_{ab},0}^{(4)} &= \frac{r_c^4}{16} \partial^2 \partial^2 ( m G^{-1} )_{ab} 
			\end{align}
			where $\partial^2=\partial_a\partial^a$. The next-to-leading coefficient $C_{I,4}^{(4)}$, however, cannot be fixed, since they give the $1/r^4$ corrections and are thus related to the solution of an homogeneous equation of the type \eqref{fluc_eq}. To determine this correction, one should in principle find the exact solution satisfying the required boundary conditions.

\subsection{Reading off the energy momentum tensor}
\label{sec:reading_EM}

In this section we describe how to read off the energy-momentum tensor of the D3-brane configuration with a slowly varying $\CF_{ab}$ field. To have a intuitive physical picture of this we consider it in ten dimensions, rather than in the equivalent five-dimensional setup of Sec.~\ref{sec:5Dsetup}. 

Consider thus a general D3-brane configuration with slowly varying $\CF_{ab}$ as described in Sec.~\ref{sec:setup} in the regime $r_c \gg l_s$. As in Sec.~\ref{sec:setup}, $R$ denotes the minimal length scale of the variation of $\CF_{ab}$ over the world-volume. The ten-dimensional metric for the full solution can be put in the form \eqref{10D_ansatz1}-\eqref{10D_ansatz2}. We require that the solution asymptotes to ten-dimensional flat space for $r\rightarrow \infty$. However, to find the metric for $r \gg R$ one needs to know $\CF_{ab}$ over a patch of size much larger than $R$. Hence for $r \gg R$ one cannot use the zeroth order solution \eqref{Sol0} as a good approximation to the full metric since one needs to go beyond the perturbative expansion in higher-derivative terms. 

Instead in the region $r_c \ll r \ll R$ one is sufficiently far away from the D3-brane to define an asymptotic region with linearized gravity, and one can use the zeroth order solution \eqref{Sol0} as a good leading order approximation to the full solution. Thus, this is the asymptotic region in which we shall read off the energy-momentum tensor of the D3-brane. 

\subsubsection*{Solution for weak gravitational field around 3-brane source}

In terms of the metric ansatz \eqref{10D_ansatz1}-\eqref{10D_ansatz2} we write the weak gravity field as
\begin{equation}
h_{ab} = \eta_{ab} + \bar{h}_{ab} \spa \eta = \bar{\eta} 
\end{equation}
where $|\bar{h}_{ab}|\ll 1$ and $|\bar{\eta}|\ll 1$. We make the gauge choice
\begin{equation}
\eta^{ab} \bar{h}_{ab} = - 4 \bar{\eta} \spa \partial^b \bar{h}_{ba} = 0
\end{equation}
In this gauge the linearized Einstein equations in the region $r_c \ll r \ll R$ become $[ \partial_r^2 + (5/r) \partial_r + \partial^c \partial_c ] \bar{h}_{ab} = 0$. If we imagine an infinitely thin 3-brane sitting at $r=0$ with energy-momentum tensor $\tau_{ab}(x^c)$ on the brane, then the linearized metric around the brane obeys the equation
\begin{equation}
\label{linEEs}
\Big[  \partial_r^2 +
\frac{5}{r} \partial_r  +  \partial^c \partial_c \Big]  \bar{h}_{ab}    = - 16 \pi G \, \delta^6(r) \tau_{ab} 
\end{equation}
where $G$ is the ten-dimensional Newtons constant given by $16\pi G = 2\pi / T_{\rm D3}^2$. Assume now $\tau_{ab}$ varies slowly along the world-volume directions and let $R$ denote the minimal length scale of that variation. Thus we expand $\tau_{ab}$ in higher-derivative corrections
\begin{equation}
\tau_{ab} = \tau_{ab}^{(0)} +  \tau_{ab}^{(2)} +\tau_{ab}^{(4)} + \CO(\partial^6)
\end{equation}
where $\tau_{ab}^{(n)}$ means that we include $n$ world-volume derivatives. Note that we assume that there can only be corrections with an even number of derivatives. Correspondingly, in the region $r \ll R$, we can also expand $\bar{h}_{ab}$ in powers of the higher-derivative contributions
\begin{equation}
\bar{h}_{ab} = \bar{h}_{ab}^{(0)} +  \bar{h}_{ab}^{(2)} +\bar{h}_{ab}^{(4)} + \CO(\partial^6)
\end{equation}
where $\bar{h}_{ab}^{(n)}$ means that we include $n$ world-volume derivatives. Inserting now these two expansions in \eqref{linEEs} we find
\begin{equation}
\begin{array}{c}\ds
\Big[  \partial_r^2 +
\frac{5}{r} \partial_r \Big] h^{(0)}_{ab} = - 16\pi G\, \delta^6(r) \tau^{(0)}_{ab}
\\[4mm] \ds
\Big[  \partial_r^2 +
\frac{5}{r} \partial_r \Big] h^{(2)}_{ab} + \partial^c \partial_c h^{(0)}_{ab} = - 16\pi G\, \delta^6(r) \tau^{(2)}_{ab}
\\[4mm] \ds
\Big[  \partial_r^2 +
\frac{5}{r} \partial_r \Big] h^{(4)}_{ab} + \partial^c \partial_c h^{(2)}_{ab} = - 16\pi G\, \delta^6(r) \tau^{(4)}_{ab}
\end{array}
\end{equation}
and so on, all in the region $r \ll R$. We can straightforwardly solve these linearized Einstein equations in the region  $r\ll R$ as
\begin{equation}
\label{solution_hbar}
\begin{array}{c} \ds
\bar{h}^{(0)}_{ab} =  \frac{r_c^4}{r^4} \frac{ \tau^{(0)}_{ab}}{NT_{\rm D3}}
\\[4mm] \ds
\bar{h}^{(2)}_{ab} =  \frac{r_c^4}{r^4} \frac{ \tau^{(2)}_{ab}}{NT_{\rm D3}} + \frac{r_c^4}{4 r^2} \frac{\partial^c \partial_c \tau^{(0)}_{ab}}{N T_{\rm D3}} 
\\[4mm] \ds
\bar{h}^{(4)}_{ab} =  \frac{r_c^4}{r^4} \frac{ \tau^{(4)}_{ab}}{NT_{\rm D3}} + \frac{r_c^4}{4 r^2} \frac{\partial^c \partial_c \tau^{(2)}_{ab}}{N T_{\rm D3}} - \frac{r_c^4}{16} \log r  \, \frac{\partial^c \partial_c \partial^d \partial_d \tau^{(0)}_{ab}}{N T_{\rm D3}}
\end{array}
\end{equation}
where we used $\Omega_5 = \pi^3$ and the definition of $r_c$ in \eqref{charge_radius}.

\subsubsection*{Reading off energy-momentum tensor for D3-brane with slowly varying $\CF_{ab}$ field}

Considering the zeroth-order solution for the metric \eqref{Sol0} we see that the leading correction to the flat space metric is
\begin{equation}
\bar{h}^{(0)}_{ab} = -  \frac{r_c^4}{r^4} m (G^{-1})_{ab}
\end{equation}
Hence according to \eqref{solution_hbar} we see that
\begin{equation}
\tau^{(0)}_{ab} = - N T_{\rm D3} m (G^{-1})_{ab}
\end{equation}
that we recognise as the EM-tensor \eqref{BI_EM2} for the leading order Born-Infeld Lagrangian \eqref{BI_LN} at strong coupling $g_s N \gg 1$. 

Considering the two-derivative correction to the gravitational solution, we see from the fact that the correction is of the form \eqref{inhomosol} with the leading coefficients given by \eqref{inhomosol_leading} that 
\begin{equation}
\bar{h}^{(2)}_{ab} = -  \frac{r_c^4}{4r^2} \partial^c \partial_c ( m G^{-1} )_{ab} 
\end{equation}
This is consistent with \eqref{solution_hbar} provided we have
\begin{equation}
\tau^{(2)}_{ab} = 0
\end{equation}
confirming what we already concluded above, that the two-derivative correction terms to the Born-Infeld action 
at strong coupling $g_s N \gg 1$ vanish. 

Finally, for the four-derivative correction to the gravitational solution we see from Eqs.~\eqref{YI4}-\eqref{YI42}
that the $\log r$ coefficient as well as the absence of a $r^{-2}$ coefficient is consistent with \eqref{solution_hbar} while the fact that we have not determined the $r^{-4}$ coefficient means that we cannot read off $\tau^{(4)}_{ab}$.

\section{Conclusions and outlook}
\label{sec:concl}

In this paper we have explored the holographic correspondence between four-dimensional Born-Infeld theory and a five-dimensional gravitational theory. This correspondence is a consequence of the open/closed string duality  \cite{Polchinski:1995mt}. There are several points to make.
\begin{itemize}
\item
The correspondence resembles the fluid/gravity correspondence  \cite{Bhattacharyya:2008jc,Hubeny:2011hd}. In particular we have the same feature that at each order in the number of world-volume derivatives, part of the Einstein equations are constraints on the variation of the collective coordinates, being the field $\CF_{ab}$, while the remaining Einstein equations should be solved to obtain the corrected gravitational solution. 
\item
One can regard the method used on the gravitational side to obtain the higher-derivative corrections as an application of the blackfold approach \cite{Emparan:2009cs,Emparan:2009at}. In particular, when reading off the energy-momentum tensor of the brane we are integrating out the short-wavelength degrees of freedom and giving a long-wavelength description for the brane in terms of collective coordinates. In this case the collective coordinates are the components of the field $\CF_{ab}$. 
\item
The Born-Infeld/gravity correspondence is a holographic correspondence for which the gravitational side is asymptotically flat. Therefore, one could potentially use it to learn about holography in flat space. While several works have investigated holography in flat space (see for example \cite{Barnich:2009se,Strominger:2013jfa} and references therein) it is a subject that still is very much in its infancy. For instance, it would be interesting to reach a deeper understanding of our prescription of how to read off the energy momentum tensor of the gravitational solution from a holographic point of view. 
\item
The open and closed string sectors are not decoupled in general. Despite this we are considering a correspondence between two effective theories in this paper, one coming from the open string sector and the other from the closed string sector. The key to understand this point seems to be the correct application of the open/closed string duality, as explained in Section~\ref{sec:OCduality}. Here it is crucial to consider the $N \gg 1$ and $g_s \ll 1$ regime which enables one to ignore string diagrams with handles. This distinguishes contributions that arise from interactions with the brane from interactions with the bulk. 
\end{itemize}

Another important question is regarding the supersymmetry of the Born-Infeld action. The Born-Infeld action has a supersymmetric completion, also when including $\alpha'$ corrections. This could suggest that the higher-derivative terms are protected by supersymmetry, also when including the planar $g_s N$ corrections. In this paper we have seen a strong hint of this, in that we have the same action at leading order at weak and strong $g_s N$ coupling, and that we found that the two-derivative corrections are absent both at weak and strong $g_s N$ coupling. On the other hand, while the derivative corrections at weak coupling come with powers of $\alpha'$, the derivative corrections at strong coupling come with powers of $\sqrt{g_s N} \alpha'$. Hence, for the four-derivative terms one could at most hope for getting the same answer at strong coupling as for weak coupling up to an overall factor depending on $g_s N$. An interesting way to illuminate these questions would be to look for supersymmetric configurations in our setup with a varying $\CF_{ab}$ field since that might be exactly solvable on both sides of the correspondence. 

We have found the new solution \eqref{Sol0} of type IIB supergravity for $N$ D3-branes with a generic constant $\CF_{ab}$ field. In section \ref{sec:duality_chain} we established a chain of string dualities and rotations that generates this solution. It would be interesting to establish what amount of supersymmetry this solution has, particularly when the D3-brane world-volume is compactified on $T^3$. 

One of the most interesting future directions would be to consider what happens in the $l_s \rightarrow 0$ decoupling limit with $\CF_{ab}$ held fixed in the limit. We established in \cite{Grignani:2013ewa} that if we take this limit for a constant $\CF_{ab}$ field then we get the AdS/CFT correspondence in the Poincar\' e patch with the world-volume metric $g_{ab} = \eta^{cd}(\eta_{ac} + \CF_{ac})(\eta_{bd}+\CF_{bd}) $, Yang-Mills coupling $\gym^2 = 4\pi g_s m$ and theta angle $\theta = 2\pi \tr ( \CF \, \hod \CF ) / ( 4 m g_s)$. This suggests that when taking the decoupling limit of the setup of Section \ref{sec:setup} with $N$ D3-branes in the background of a slowly varying $\CF_{ab}$ field, then one obtains a new duality between $SU(N)$ $\CN=4$ SYM theory on a curved space  with varying Yang-Mills coupling and theta angle, and a gravitational background that locally is approximately $\ads_5\times S^5$ in the Poincar\' e patch, but over large distances - the distances over which $\CF_{ab}$ vary - is modified. 

Finally it would be worth to study the Born-Infeld/gravity correspondence at non-zero temperature, repeating the analysis 
presented in this paper for a system of $N$ coincident non-extremal D3-branes. We know that when a constant $\CF_{ab}$ field 
is turned on on the brane the leading $T^4$ terms of the free energies at weak 
and strong coupling only differ by a 3/4 factor \cite{Grignani:2013ewa}, 
just like in absence of $\CF_{ab}$ \cite{Gubser:1996de}. This result holds exactly because, when $\CF_{ab}$
is constant, the decoupling limit yield again the AdS/CFT correspondence in the Poincar\' e patch, 
only in a different coordinate system. It would be then very interesting to investigate what happens, also thermodynamically,
when $\CF_{ab}$ is allowed to vary slowly on the brane world-volume.

\section*{Acknowledgments}

We thank Jan de Boer, Jelle Hartong and Niels Obers for useful discussions. TH acknowledge support from the European Union Marie-Curie-CIG grant ``Quantum Mechanical Nature of Black Holes".


\small

\addcontentsline{toc}{section}{References}


\providecommand{\href}[2]{#2}\begingroup\raggedright\endgroup

\end{document}